\documentclass[conference, letterpaper]{IEEEtran}
\IEEEoverridecommandlockouts
\usepackage{cite}
\usepackage{amsmath,amssymb,amsfonts}
\usepackage{algorithmic}
\usepackage{graphicx}
\usepackage{textcomp}
\usepackage{xcolor}
\usepackage{enumitem}
\def\BibTeX{{\rm B\kern-.05em{\sc i\kern-.025em b}\kern-.08em
    T\kern-.1667em\lower.7ex\hbox{E}\kern-.125emX}}
\DeclareMathOperator\arctanh{arctanh}

\begin{document}

\title{Evaluation of Sampling Algorithms for a Pairwise Subjective Assessment Methodology\\

}

\author{
    \IEEEauthorblockN{Shima Mohammadi\IEEEauthorrefmark{1}, Jo\~{a}o Ascenso\IEEEauthorrefmark{2}}
    \IEEEauthorblockA{\IEEEauthorrefmark{1}\IEEEauthorrefmark{2}Instituto Superior Técnico - Instituto de Telecomunicações}
    \IEEEauthorblockA{\IEEEauthorrefmark{1}shima.mohammadi@lx.it.pt\IEEEauthorrefmark{2}joao.ascenso@lx.it.pt}
}

\maketitle

\begin{abstract}
Subjective assessment tests are often employed to evaluate image processing systems, notably image and video compression, super-resolution among others and have been used as an indisputable way to provide evidence of the performance of an algorithm or system. While several methodologies can be used in a subjective quality assessment test, pairwise comparison tests are nowadays attracting a lot of attention due to their accuracy and simplicity. However, the number of comparisons in a pairwise comparison test increases quadratically with the number of stimuli and thus often leads to very long tests, which is impractical for many cases. However, not all the pairs contribute equally to the final score and thus, it is possible to reduce the number of comparisons without degrading the final accuracy. To do so, pairwise sampling methods are often used to select the pairs which provide more information about the quality of each stimuli. In this paper, a reliable and much-needed evaluation procedure is proposed and used for already available methods in the literature, especially considering the case of subjective evaluation of image and video codecs. The results indicate that an appropriate selection of the pairs allows to achieve very reliable scores while requiring the comparison of a much lower number of pairs.
\end{abstract}

\begin{IEEEkeywords}
Image quality assessment, Subjective assessment, Pairwise comparison, Pairwise sampling
\end{IEEEkeywords}
\vspace{-7pt}
\section{Introduction}

Image and video quality assessment is an essential tool for many computer graphics and multimedia applications with the ambition to improve users’ perceived Quality of Experience. Image and video quality assessment falls into two categories: objective quality metrics and subjective assessment. In full-reference objective quality metrics, the image is fed into a mathematical model of human perception that expresses the distortion (or quality degradation) between a reference and a distorted image. Although, objective quality metrics are automatic and easy to perform, it still does not eliminate the need for subjective assessment, which is considered a more accurate and reliable way to measure the perceived quality of some system. Moreover, scores obtained from subjective assessment tests are often used as the ground truth in the design and evaluation of objective quality metrics, which is rather needed when new image processing systems are deployed, e.g. benchmarking an emerging image compression solution.

To obtain subjective scores, several subjective assessment methodologies are suggested in standards or guidelines \cite{BT500-19}\cite{P910-21}\cite{JPEG-AIC-Part2}. Nowadays, the most popular methods rate the content, using a single or double stimulus methodology, depending on how the stimuli are presented. Typically, subjects are asked to score the stimuli using either a numerical (categorical) or continuous quality scale. Absolute category rating (ACR) \cite{P910-21} for single-stimulus and double-stimulus impairment scale or double-stimulus continuous quality scale \cite{BT500-19} are the most popular subjective methodologies.

\begin{figure}[htbp]
\vspace{-12pt}
\centerline{
\begin{tabular}{@{}c@{}}
    {\includegraphics[scale=0.2]{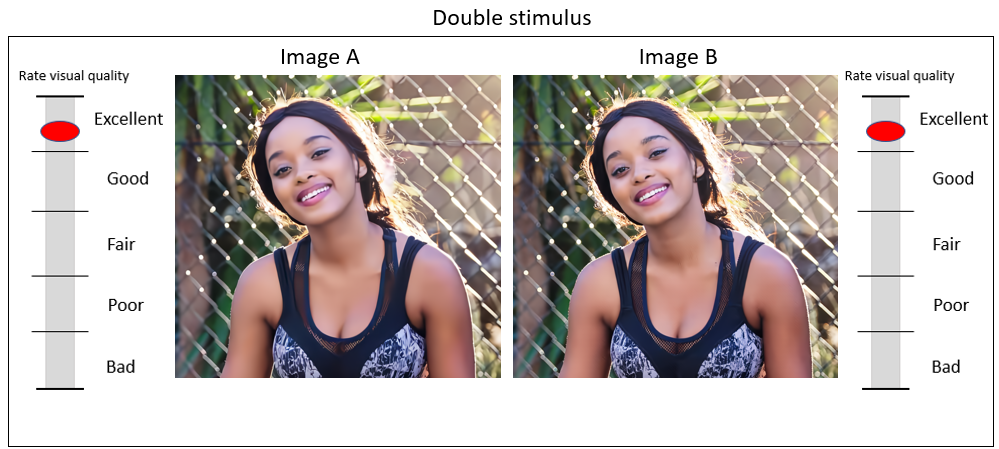}} \\
  \end{tabular}}
\centerline{
\begin{tabular}{@{}c@{}}
    {\includegraphics[scale=0.2]{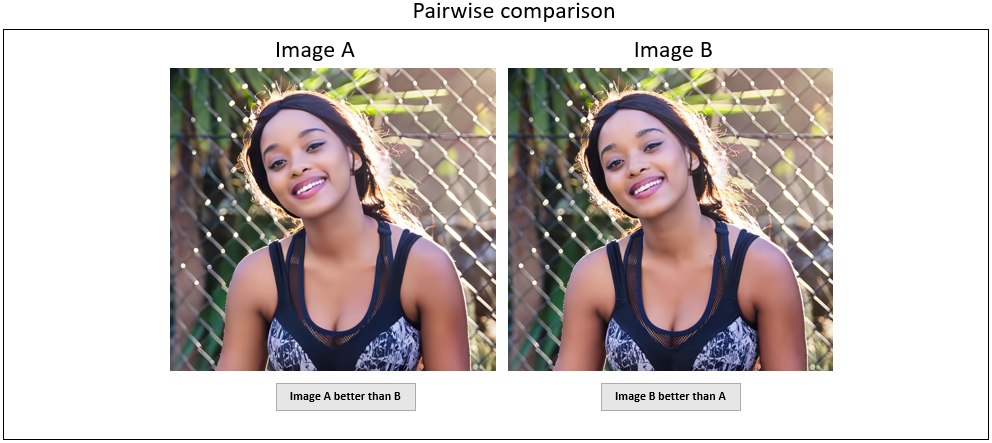}} \\
  \end{tabular}}
\caption{Subjective image quality assessment methodologies: double-stimulus continuous quality scale (top) and pairwise comparison (bottom)}  
\label{ACR_vs_PC}
\vspace{-9pt}
\end{figure}

However, several studies \cite{Pinson2003}\cite{Tominaga2010}\cite{Mantiuk2012} have shown that pairwise comparison (PC) is more reliable and accurate than other methods, such as single and double-stimulus methods, mainly because the evaluation is simpler (and therefore faster) and small differences between stimuli can be more accurately measured. For example, this type of subjective methodology is very promising to assess image coding solutions, providing discriminative scores between decoded images from high quality to near-visually lossless quality. Actually, the JPEG Committee has launched recently a new activity on Assessment of Image Coding (AIC), also referred as JPEG AIC to develop a new standard (AIC-3) \cite{JPEGAIC-3b}\emph{ focusing on the methodologies for assessment of images with the quality levels in between the range where ITU-Rec. BT.500 \cite{BT500-19} is suitable and the range where AIC-2 \cite{JPEG-AIC-Part2} is suitable}.  Fig.\ref{ACR_vs_PC} illustrates the type of decisions subjects have to make with a double-stimulus continuous quality scale methodology and with a pairwise comparison methodology.

The PC subjective assessment methodology consists of the comparison of two images or videos (often, side by side) to evaluate which one is preferred; it can also be used to assess if the two images are identical in quality \cite{Lee2011} \cite{ITU910}. Therefore, PC tests can be understood as a ranking (and not rating) method, where subjects decide which images have the highest quality. This binary judgment has several advantages, namely not requiring training associated with the meaning of the quality scale (also avoiding any misunderstanding) and thus, more suitable for non-expert viewers. In addition, it is more robust to changes in viewing conditions, since subjects judgment does not change much, e.g., when the viewing distance or screen size changes. This is rather suitable for crowdsourcing tests, requiring no (or little) subject training while providing reliable results.

A complete PC design is when all the possible pair combinations between all stimuli are used. Complete design requires $n\times(n-1)/2$ pairs to be compared where $n$ is the number of stimuli. This means that the number of pairs to be compared increases quadratically ($\mathcal{O}(n^2)$) and thus the length of the test is impractical for a wide range of problems, especially when a large number of contents or degradations have to be considered. To overcome this problem, an incomplete design where only a subset of the pairs is compared can be used. This comes from the observation that not all comparisons provide useful information for the final ranking or score, e.g. comparison of pairs of images which have very different qualities.

In image and video quality assessment studies employing a PC methodology, it is often used a complete design \cite{Zhang2019}\cite{Jiang2022} or an incomplete design where the \textit{Swiss-design} system tournament style \cite{ponomarenko2015} performs the selection (or sampling) of the image pairs \cite{Ko2020}\cite{Mikhailiuk2019}. However, there are many other selection criteria, based on heuristics \cite{Reinhard2019} or some information-theoretic criteria \cite{Ye2014}. Therefore, the main paper objectives are: 
\begin{enumerate}
 \item Propose an evaluation procedure, which was never clearly defined in previous work \cite{chen2013}\cite{li2018} for the evaluation of pairwise reduction algorithms, considering the characteristics of image and video quality subjective assessment experiments. 
 \item Benchmark selected state-of-the-art pairwise sampling methods using the proposed evaluation method, providing a much-needed reference for those interested in performing pairwise comparison tests in the future.
\end{enumerate}
\vspace{-7pt}
\section{Short Review of Pairwise Sampling Methods}\label{sec-review}

Pairwise comparison tests require the subjects to select one out of two images in the pair. The judgments made by the subjects are used to infer the latent scores, typically using psychometric scaling methods, such as the Thurstone’s \cite{Thurstone1927} or Bradley-Terry’s (BT) models \cite{Bradley1952}. The methods for the selection of pairs can be classified into three classes: random, sorting, and active sampling. Moreover, the methods can also be sequential, where the selection of the next pair can only be performed after the judgment of the previous pair is received, and batch (or parallel), where a set of pairs is generated and judgments are jointly received (in a synchronous way). Batch methods are preferred in online crowdsourcing subjective tests, where a set of subjects can perform the test in parallel. In the following, relevant state-of-the-art pairwise sampling methods are described; note that the methods selected for performance evaluation (Section \ref{sec-performance}) are highlighted in italic.
\begin{enumerate}[leftmargin=.2in]
\item \textbf{Random}: The simplest way to select pairs of images is based on random sampling which does not exploit information from past comparisons, and each pair of images have the same probability of being selected. In \textit{HR-random} \cite{Xu2012a}, the sampling strategy is based on Erdos-Renyi random graphs, and the Hodge theory is applied to convert incomplete and imbalanced data to scaled ratings.

\item \textbf{Sorting}: Sorting based methods use a ranking procedure based on some sorting method (e.g. insertion sort), in a way that pairs of images with similar quality are more often selected than pairs of images with different quality. 
\begin{enumerate}[leftmargin=.2in]
\item \textit{Swiss-design} \cite{ponomarenko2015}: The most often used solution is the \textit{Swiss-design} Tournament system. This approach has two rounds. First, random pairs are selected for comparison then, in the second round, images with similar quality are selected in the creation of new pairs. 
\item Quicksort \cite{Quicksort}: Quicksort algorithm is applied to select the pairs of images and reduce the number of comparisons.
\end{enumerate}

\begin{figure*}[ht]
\vspace{-14pt}
\setlength\abovecaptionskip{0pt}
\setlength\belowcaptionskip{0pt}
\centerline{
\begin{tabular}{@{}ccccc@{}}
    {\includegraphics[scale=0.6]{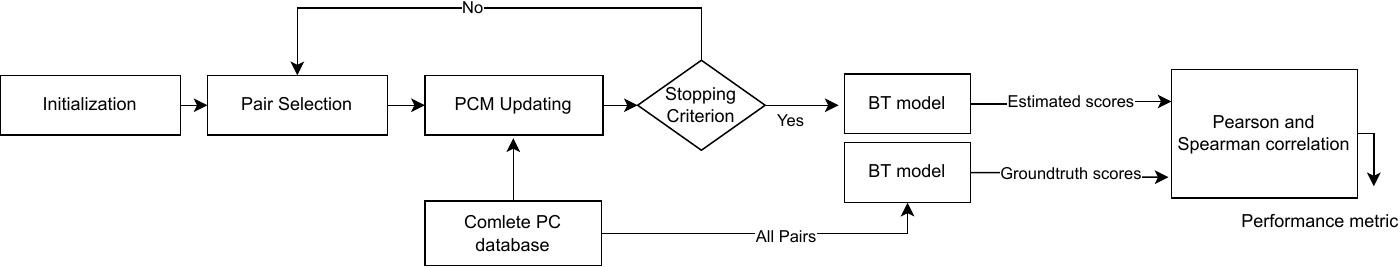}} \\
  \end{tabular}}
\caption{Proposed evaluation procedure flowchart.}
\label{flowchart}
\vspace{-9pt}
\end{figure*}
\item \textbf{Active}: In this approach the outcome of previous judgments determine the selection of the images in the creation of the next pairs. The active sampling strategy aims to select only the most informative pairs for comparison by exploiting information from the prior and posterior distribution of the latent quality scores (usually obtained from the BT model). A pair is selected according to a utility function, such as Kullback-Leibler (KL) divergence between the prior distribution and posterior distribution assuming any possible outcome. The following active sampling methods are described next:

\begin{enumerate}[leftmargin=.2in]
\item \textit{Active HodgeRank (HR-active)}\cite{Xu2018}: The selection of pairs is made in the framework of the HodgeRank model and as a Bayesian information maximization procedure to select the pairs with the largest information gain from the prior and posterior distributions of quality scores. 
\item \textit{Crowdsourcing Bradley-Terry (Crowd-BT)} \cite{chen2013}: The reliability of the subjects's decision is taken into account, namely giving less weight to subjects that make random or wrong judgments.
\item \textit{Hybrid minimum spanning tree (Hybrid-MST)}\cite{li2018}: Hybrid approach based on a minimum spanning tree and maximum expected information gain, which is defined as the KL divergence between the prior distribution and the posterior distribution when all possibles outcomes are considered (one stimuli is preferred over the other and the opposite). 
\item \textit{Active Sampling for Pairwise Comparisons (ASAP)} \cite{asap}: \textit{ASAP} represents an improvement to \textit{Hybrid-MST} where the posterior distribution is updated using the entire set of comparisons, and the utility function is computed only for a subset of the pairs.
\item Hybrid subjective test methodology (HY-Subj) \cite{Peng2014}\cite{Linq2020}: rating and ranking approaches are combined through a unified probabilistic model and an active sampling method. The idea is to use a rating method to provide a coarse estimate of the underlying quality followed by a PC test to provide fine discrimination for images with similar quality.
\item Active Binary Tree (ABT) \cite{Silverstein2001binarytree}: This approach relies on the assumption that pairs which are more close in quality should be more often compared than distant pairs. It was shown that by sorting the data with a pilot study before the pairwise comparison test, fewer pairs are needed to reliably rank the data. Moreover, the proposed algorithm relies on the construction of a binary tree which helps to decide which images should be compared.
\item Reliable informativeness PC (RI-PC)\cite{Zhiwei2017}: This work accounts for pairs of images that are difficult to rank, i.e. where the difference between two images is quite small, and thus the decision is not easy and a matter of ambiguous opinion. The sampling strategy is therefore enhanced by including both the informativeness of pairs (information that a pair can contribute to obtain accurate scores) and the effort spent on an image pair due to its ambiguity.
\item Particle Filtering PC \cite{storrs2018}: An adaptive pairwise comparison method based on particle filtering method to evaluate subjective test on mobile devices is proposed. The pairs of images shown to the subject were determined by an active learning procedure, based on the subjects's responses on previous judgements.
\end{enumerate}
\end{enumerate}

\vspace{-7pt}
\section{Proposed Evaluation Procedure}\label{sec-propos_eval}

The key data structure of a PC subjective assessment test is the PC Matrix ($PCM$) which is calculated based on the judgments made by subjects. $PCM[i,j]$ represents the number of times image $i$ is preferred over image $j$ and can be converted (e.g. using a BT model) to psychometric scores that define the quality of each degraded image. The proposed evaluation procedure (as shown in Fig.\ref{flowchart}) can be defined as follows:
\begin{enumerate}[leftmargin=.2in]
\item \textbf{Initialization}: The $PCM[i,j]$ is initialized with all $1$s, this means that all pairs have been evaluated and have the same quality, which avoids having pairs for which no comparison is performed. 
\item \textbf{Pair Selection}: Pairs of images are selected according to the pairwise sampling algorithm under evaluation. While for some algorithms one pair is selected each time, for others like \textit{Hybrid-MST} multiple pairs can be selected in this step (batch mode facility).
\item \textbf{$\boldsymbol{PCM}$ Updating}: Using some already available dataset for which pairwise comparisons are available for all pairs, a judgment is selected for each selected pair of the previous step; this judgment is selected randomly and represents the assessment made by some subject. Then, the $PCM$ is updated based on the judgments obtained.
\item \textbf{Stopping Criterion}: Steps 2 and 3 of the algorithm are repeated until the stopping criterion is met. In this case, this is the maximum number of pairs that can be selected (budget), which directly defines the length of the subjective test. Note also that the updated $PCM$ in the previous step can be used for the selection of the next pairs (Step 2) in the next iteration.  
\end{enumerate}

After the pair budget has been reached, an incomplete PC test has been simulated and after, the estimated psychometric scores are calculated with the BT model from the obtained $PCM$. The ground truth scores are obtained in a similar way but using all the possible pair comparisons. Naturally, the datasets to be used must have been obtained using a complete design, although they can be imbalanced, i.e. the number of votes for each stimuli can be different. The performance of some pairwise sampling algorithm is analysed using the Pearson linear correlation coefficient (PLCC) and Spearman rank order correlation coefficient (SROCC) between the ground truth (complete design) and simulated scores (incomplete design), for some algorithm and pair budget.

\vspace{-3pt}
\section{Performance Evaluation}\label{sec-performance}
\begin{figure*}[htbp]
\vspace{-14pt}
\centerline{
\begin{tabular}{@{}c@{}}
\vspace{-4pt}
    {\includegraphics[scale=0.32]{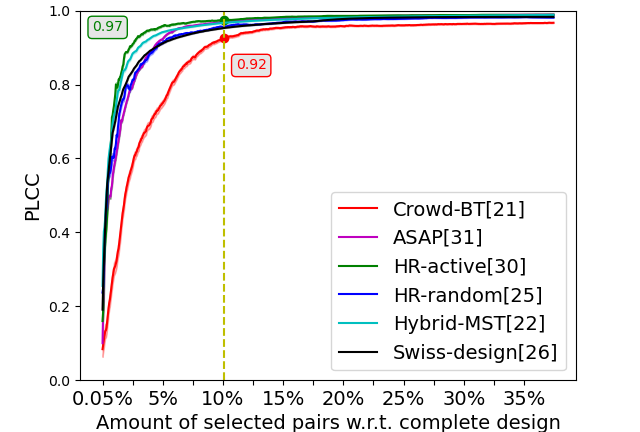}} \\ 
    \scriptsize(a) Average PLCC for the IQA dataset
  \end{tabular}
  \begin{tabular}{@{}c@{}}
  \vspace{-4pt}
    {\includegraphics[scale=0.32]{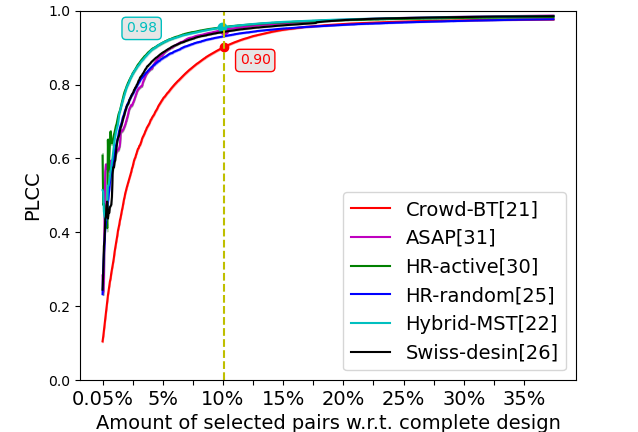}} \\ 
    \scriptsize(b) Average PLCC for the VQA dataset
  \end{tabular}
  \begin{tabular}{@{}c@{}}
  \vspace{-4pt}
    {\includegraphics[scale=0.32]{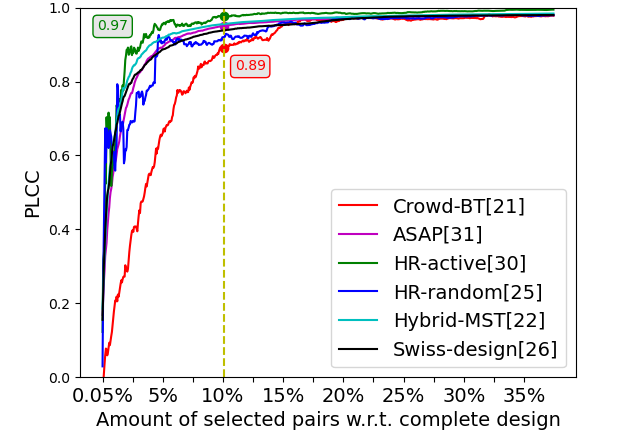}} \\ 
    \scriptsize(c) Average PLCC for the synthetic dataset
  \end{tabular}
  }
  \vspace{-2pt}
  \centerline{
     \begin{tabular}{@{}c@{}}
     \vspace{-4pt}
    {\includegraphics[scale=0.32]{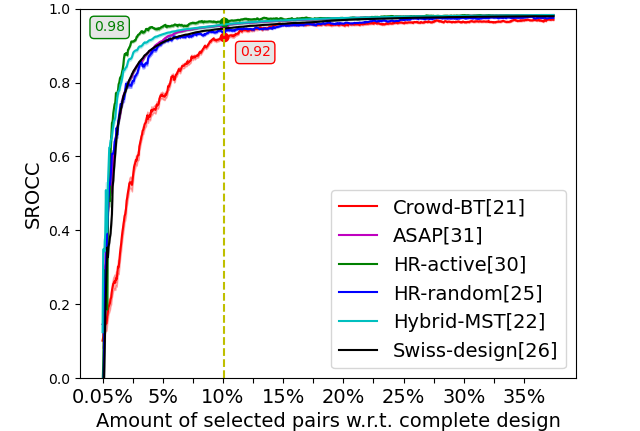}} \\ 
    \scriptsize (d) Average SROCC for the IQA dataset
   
  \end{tabular}

  \begin{tabular}{@{}c@{}}
  \vspace{-4pt}
    {\includegraphics[scale=0.32]{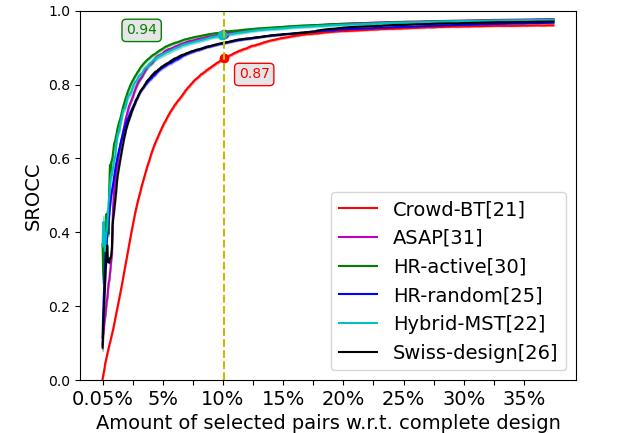}} \\ 
    \scriptsize (e) Average SROCC for the VQA dataset
   
  \end{tabular}
  \begin{tabular}{@{}c@{}}
  \vspace{-4pt}
    {\includegraphics[scale=0.32]{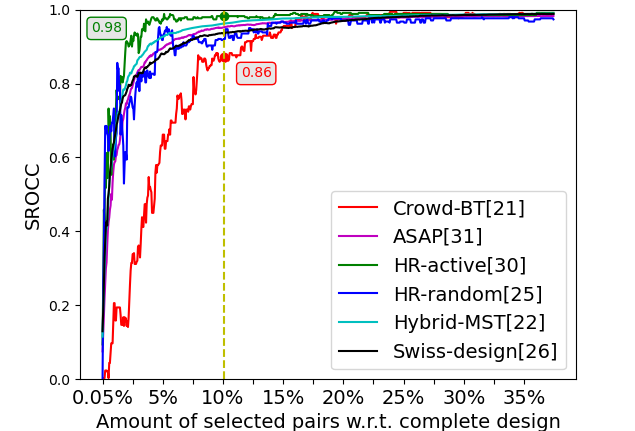}} \\ 
    \scriptsize (f) Average SROCC for the synthetic dataset
  \end{tabular}
  }
\caption{Performance evaluation for all the benchmark pairwise sampling methods using data obtained from three selected datasets.}
\label{fig_results}
\vspace{-9pt}
\end{figure*}
The selected state-of-the-art pairwise sampling algorithms are evaluated using two different types of datasets. The first dataset was artificially obtained while the remaining two datasets were obtained by performing subjective experiments on images and videos. While in rating methodologies, some subjective evaluation parameters need to be controlled (such as display size and resolution, viewing distance), pairwise comparison tests are rather robust to these changes and thus no restrictions are typically enforced.
\vspace{-7pt}
\subsection{Synthetic Dataset}
A synthetic dataset was generated for one reference image with 16 degraded versions. For each degraded image, a MOS score and corresponding standard deviation was obtained by sampling a uniform distribution with values between 1 to 5 and 0 to 0.7, respectively (according to the distribution of the subjective scores obtained in the large scale subjective test of \cite{VQEG-report}). Thus, the quality of every degraded image is represented with a Normal distribution, with parameters $\mu$ (=MOS) and $\sigma$. For each subject, opinion scores can be obtained by sampling this distribution. Thus, for every pair $(i,j)$, the score of images $i$ and $j$ is obtained by sampling the corresponding quality distribution and the judgment of some subject is obtained with a simple rule: if the score for $i$ is higher than the score of $j$, $i$ is better than $j$. Furthermore, to consider the real scenario of having unreliable subjects which gives false ranking, the preference is inverted with 10\% percent probability. In \cite{TID2013} a lab controlled subjective assessment experiment was performed, 2\% of the votes were identified as outliers and thus 10\% is used here since the target is to simulate a crowdsourcing approach, where more outliers are typically present due to the uncontrolled nature of the test. 

\subsection{Real Datasets}
The two selected real datasets are described next: 
\begin{itemize}[leftmargin=.2in]
\item Image quality assessment dataset (IQA) \cite{Xu2012b} is a complete but imbalanced dataset, where a subjective test via a crowdsourcing platform was performed on images of LIVE \cite{LIVE-IQA} and IVC dataset \cite{Callet2005}. This data set contains 15 reference images each with 16 distorted versions. 
\item Video quality assessment dataset (VQA)\cite{Qianqian2011} by assuming 32 subjects is a complete and balanced dataset, where a subjective test via a crowdsourcing platform was performed on videos of LIVE dataset \cite{LIVE_VQA}. Each video in this dataset contains 10 distorted versions.
\end{itemize}

\vspace{-7pt}
\subsection{Experimental Results}
The evaluation procedure is conducted on the aforementioned datasets. Each selected algorithm is repeated one hundred times for each reference, to have an unbiased test and the average PLCC and SROCC is computed by averaging over all repetitions. The average PLCC and SROCC over all references is computed for the results shown here. Also, note that the selection of the judgments is made randomly, according to the procedure proposed in Section \ref{sec-propos_eval}. 

The results obtained are shown in Fig. \ref{fig_results}. In these figures the horizontal axis represent the percentage of the pairs selected with respect to having a complete design test for 15 subjects, where 15 is the minimum number of subjects based on the ITU-T P.910 \cite{P910-21} recommendation.

As shown in Fig. \ref{fig_results}, the correlation of the active sampling methods is greater than 0.9 for only 10\% of all the possible pairs (i.e. with respect to the complete design) and can be as high as 0.97 and as low as 0.86; for 35\% of the pairs, most of the methods reach a correlation close to $1$. Thus, only a short amount of comparisons need to be performed, while maintaining all the advantages of the pairwise test over other rating methods. Note pairwise comparison test are typically more accurate than other rating methodologies such as double stimulus test.

\begin{figure}[htbp]
\vspace{-14pt}
\centerline{
\begin{tabular}{@{}c@{}}
\vspace{-4pt}
    {\includegraphics[scale=0.25]{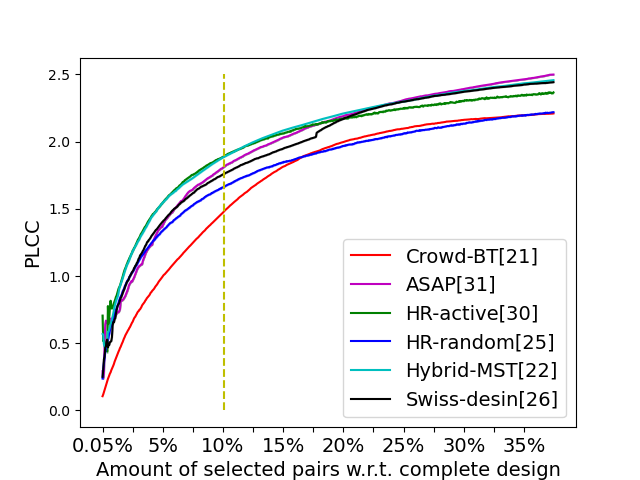}} \\
   
  \end{tabular}
  
  \vspace{-2pt}

\begin{tabular}{@{}c@{}}

    {\includegraphics[scale=0.25]{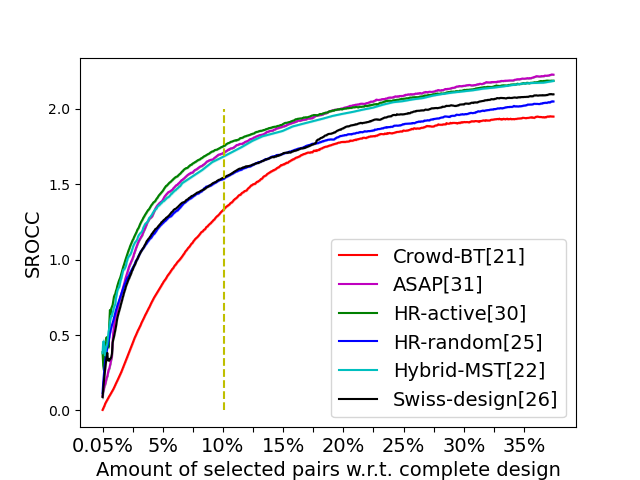}} \\
  \end{tabular}}
\caption{Performance evaluation for all the benchmark pairwise sampling methods after fisher transformation for VQA dataset}  
\label{fisher_tr}
\vspace{-5pt}
\end{figure}

In addition, Fig.\ref{fisher_tr} shows the results after the fisher transformation $y'=\arctanh(y)$ is applied to scale the PLCC values and thus show better the difference between curves.The active sampling methods \textit{HR-active}, \textit{ASAP} and \textit{Hybrid-MST} have higher performance than random sampling \textit{HR-random}, \textit{Swiss-design} and \textit{Crowd-BT}. Regarding SROCC, when the percentage of the pairs are small, \textit{HR-active} outperform the other algorithms (but not much for PLCC), while as the percentage of the pairs increases, \textit{ASAP}, and \textit{Hybrid-MST} outperform \textit{HR-active}. On the other hand, the performance of \textit{Swiss-design} although very simple, has approximately same or higher SROCC performance than random sampling approaches for low and medium percentages. However, the \textit{Swiss-design} algorithm outperform \textit{HR-active} when 20 percent of the pairs are compared in VQA dataset for the PLCC performance metric. Between the different real datasets, the performance is similar for the several algorithms, although a slightly lower performance can be observed for the VQA dataset. Moreover, it is also important to notice that the results obtained for the synthetic dataset are more noisy mainly because just one reference image was used and that HR-active clearly provides the best performance both in SROCC and PLCC.
\vspace{-6pt}
\vspace{-2pt}
\section{Conclusions}\label{sec-conclusions}
This paper targets the feasibility of pairwise comparison subjective assessment by first proposing a procedure to evaluate the performance of pairwise sampling algorithms, which are responsible to reduce the length of this type of subjective test. The proposed evaluation framework allows the assessment of current and future pairwise sampling algorithms under precise and meaningful conditions. Experimental results are obtained for three datasets (one synthetic and two real) using several state-of-the-art pairwise sampling algorithms, providing a much-needed reference in this area. From the analysis of the results, it was concluded that the active sampling methods have higher performance than other methods achieving very high correlations with just 10\% of all possible pairs.

\bibliographystyle{IEEEtran}
\bibliography{citation}

\begin{thebibliography}{10}
\providecommand{\url}[1]{#1}
\csname url@samestyle\endcsname
\providecommand{\newblock}{\relax}
\providecommand{\bibinfo}[2]{#2}
\providecommand{\BIBentrySTDinterwordspacing}{\spaceskip=0pt\relax}
\providecommand{\BIBentryALTinterwordstretchfactor}{4}
\providecommand{\BIBentryALTinterwordspacing}{\spaceskip=\fontdimen2\font plus
\BIBentryALTinterwordstretchfactor\fontdimen3\font minus
  \fontdimen4\font\relax}
\providecommand{\BIBforeignlanguage}[2]{{%
\expandafter\ifx\csname l@#1\endcsname\relax
\typeout{** WARNING: IEEEtran.bst: No hyphenation pattern has been}%
\typeout{** loaded for the language `#1'. Using the pattern for}%
\typeout{** the default language instead.}%
\else
\language=\csname l@#1\endcsname
\fi
#2}}
\providecommand{\BIBdecl}{\relax}
\BIBdecl

\bibitem{BT500-19}
{ITU-R Question 102-3/6}, ``Recommendation 500-19: Methodology for the
  subjective assessment of the quality of television pictures,'' \emph{ITU-R
  Recommendation BT}, 2019.

\bibitem{P910-21}
{ITU-T Study Group 12}, ``Subjective video quality assessment methods for
  multimedia applications,'' \emph{ITU-T Recommendation P.910}, 2021.

\bibitem{JPEG-AIC-Part2}
{ISO/IEC WG1}, ``Information technology — advanced image coding and
  evaluation — part 2: Evaluation procedure for nearly lossless coding,''
  \emph{ISO/IEC 29170-2}, 2015.

\bibitem{Pinson2003}
M.~H. Pinson and S.~Wolf, ``Comparing subjective video quality testing
  methodologies,'' in \emph{Visual Communications and Image Processing}, vol.
  5150.\hskip 1em plus 0.5em minus 0.4em\relax SPIE, 2003, pp. 573--582.

\bibitem{Tominaga2010}
T.~Tominaga, T.~Hayashi, J.~Okamoto, and A.~Takahashi, ``Performance
  comparisons of subjective quality assessment methods for mobile video,'' in
  \emph{International Workshop on Quality of Multimedia Experience}, Trondheim,
  Norway, 2010, pp. 82--87.

\bibitem{Mantiuk2012}
R.~K. Mantiuk, A.~Tomaszewska, and R.~Mantiuk, ``Comparison of four subjective
  methods for image quality assessment,'' in \emph{Computer graphics forum},
  vol.~31, no.~8.\hskip 1em plus 0.5em minus 0.4em\relax Wiley Online Library,
  2012, pp. 2478--2491.

\bibitem{JPEGAIC-3b}
{ISO/IEC JTC 1/SC29/WG1 N100163}, ``Review of the state of the art on
  subjective image quality assessment,'' in \emph{95th WG1 Meeting, Online,
  April 2022}.

\bibitem{Lee2011}
J.-S. Lee, L.~Goldmann, and T.~Ebrahimi, ``A new analysis method for paired
  comparison and its application to {3D} quality assessment,'' in \emph{ACM
  International Conference on Multimedia}, New York, NY, USA, 2011, p.
  1281–1284.

\bibitem{ITU910}
ITU-T, ``P.910, subjective video quality assessment methods for multimedia
  applications,'' in \emph{International Telecommunication Union}, 1999.

\bibitem{Zhang2019}
X.~Zhang, W.~Lin, S.~Wang, J.~Liu, S.~Ma, and W.~Gao, ``Fine-grained quality
  assessment for compressed images,'' \emph{IEEE Transactions on Image
  Processing}, vol.~28, no.~3, pp. 1163--1175, 2019.

\bibitem{Jiang2022}
Q.~Jiang, Z.~Liu, K.~Gu, F.~Shao, X.~Zhang, H.~Liu, and W.~Lin, ``Single image
  super-resolution quality assessment: A real-world dataset, subjective
  studies, and an objective metric,'' \emph{IEEE Transactions on Image
  Processing}, vol.~31, pp. 2279--2294, 2022.

\bibitem{ponomarenko2015}
N.~Ponomarenko, L.~Jin, O.~Ieremeiev, V.~Lukin, K.~Egiazarian, J.~Astola,
  B.~Vozel, K.~Chehdi, M.~Carli, F.~Battisti \emph{et~al.}, ``Image database
  tid2013: Peculiarities, results and perspectives,'' \emph{Signal processing:
  Image communication}, vol.~30, pp. 57--77, 2015.

\bibitem{Ko2020}
H.~Ko, D.~Y. Lee, S.~Cho, and A.~C. Bovik, ``Quality prediction on deep
  generative images,'' \emph{IEEE Transactions on Image Processing}, vol.~29,
  pp. 5964--5979, 2020.

\bibitem{Mikhailiuk2019}
A.~Mikhailiuk, M.~Pérez-Ortiz, and R.~Mantiuk, ``Psychometric scaling of
  {TID2013} dataset,'' in \emph{International Conference on Quality of
  Multimedia Experience}, Sardinia, Italy, 2018, pp. 1--6.

\bibitem{Reinhard2019}
R.~Heckel, N.~B. Shah, K.~Ramchandran, and M.~J. Wainwright, ``Active ranking
  from pairwise comparisons and when parametric assumptions do not help,''
  \emph{The Annals of Statistics}, vol.~47, no.~6, pp. 3099 -- 3126, 2019.

\bibitem{Ye2014}
P.~Ye and D.~Doermann, ``Active sampling for subjective image quality
  assessment,'' in \emph{IEEE Conference on Computer Vision and Pattern
  Recognition}, Ohio, USA, 2014, pp. 4249--4256.

\bibitem{chen2013}
X.~Chen, P.~N. Bennett, K.~Collins-Thompson, and E.~Horvitz, ``Pairwise ranking
  aggregation in a crowdsourced setting,'' in \emph{ACM international
  conference on Web search and data mining}, Rome, Italy, 2013, pp. 193--202.

\bibitem{li2018}
J.~Li, R.~Mantiuk, J.~Wang, S.~Ling, and P.~Le~Callet, ``{Hybrid-MST}: A hybrid
  active sampling strategy for pairwise preference aggregation,''
  \emph{Advances in neural information processing systems}, vol.~31, p.
  3479–3489, 2018.

\bibitem{Thurstone1927}
L.~L. Thurstone, ``A law of comparative judgment.'' \emph{Psychological
  review}, vol.~34, no.~4, p. 273, 1927.

\bibitem{Bradley1952}
R.~A. Bradley and M.~E. Terry, ``Rank analysis of incomplete block designs: I.
  the method of paired comparisons,'' \emph{Biometrika}, vol.~39, no. 3/4, pp.
  324--345, 1952.

\bibitem{Xu2012a}
Q.~Xu, Q.~Huang, T.~Jiang, B.~Yan, W.~Lin, and Y.~Yao, ``{HodgeRank} on random
  graphs for subjective video quality assessment,'' \emph{IEEE Transactions on
  Multimedia}, vol.~14, no.~3, pp. 844--857, 2012.

\bibitem{Quicksort}
A.~Mikhailiuk, C.~Wilmot, M.~Pérez-Ortiz, D.~Yue, and R.~Mantiuk, ``Active
  sampling for pairwise comparisons via approximate message passing and
  information gain maximization,'' 05 2020.

\bibitem{Xu2018}
Q.~Xu, J.~Xiong, X.~Chen, Q.~Huang, and Y.~Yao, ``{HodgeRank} with information
  maximization for crowdsourced pairwise ranking aggregation,'' in \emph{AAAI
  Conference on Artificial Intelligence and Innovative Applications of
  Artificial Intelligence Conference and AAAI Symposium on Educational Advances
  in Artificial Intelligence}, New Orleans, Louisiana, USA, 2018.

\bibitem{asap}
A.~Mikhailiuk, C.~Wilmot, M.~Perez-Ortiz, D.~Yue, and R.~K. Mantiuk, ``Active
  sampling for pairwise comparisons via approximate message passing and
  information gain maximization,'' in \emph{2020 25th International Conference
  on Pattern Recognition (ICPR)}, Milan, Italy, 2021, pp. 2559--2566.

\bibitem{Peng2014}
P.~Ye and D.~Doermann, ``Active sampling for subjective image quality
  assessment,'' in \emph{2014 IEEE Conference on Computer Vision and Pattern
  Recognition}, OH, USA, 2014, pp. 4249--4256.

\bibitem{Linq2020}
S.~Ling, J.~Li, A.~F. Perrin, Z.~Li, L.~Krasula, and P.~L. Callet, ``Strategy
  for boosting pair comparison and improving quality assessment accuracy,''
  \emph{arXiv preprint arXiv:2010.00370}, 2020.

\bibitem{Silverstein2001binarytree}
D.~A. Silverstein and J.~E. Farrell, ``Efficient method for paired
  comparison,'' \emph{J. Electronic Imaging}, vol.~10, pp. 394--398, 2001.

\bibitem{Zhiwei2017}
Z.~Fan, T.~Jiang, and T.~Huang, ``Active sampling exploiting reliable
  informativeness for subjective image quality assessment based on pairwise
  comparison,'' \emph{IEEE Transactions on Multimedia}, vol.~19, no.~12, pp.
  2720--2735, 2017.

\bibitem{storrs2018}
K.~Storrs, S.~Van~Leuven, S.~Kojder, L.~Theis, and F.~Huszár, ``Adaptive
  paired-comparison method for subjective video quality assessment on mobile
  devices,'' in \emph{2018 Picture Coding Symposium (PCS)}, CA, USA, 2018, pp.
  169--173.

\bibitem{VQEG-report}
\BIBentryALTinterwordspacing
VQEG, ``Report on the validation of video quality models for high definition
  video content,'' 2010. [Online]. Available: \url{\url{http://www.vqeg.org}}
\BIBentrySTDinterwordspacing

\bibitem{TID2013}
N.~Ponomarenko, L.~Jin, O.~Ieremeiev, V.~Lukin, K.~Egiazarian, J.~Astola,
  B.~Vozel, K.~Chehdi, M.~Carli, F.~Battisti \emph{et~al.}, ``Image database
  tid2013: Peculiarities, results and perspectives,'' \emph{Signal processing:
  Image communication}, vol.~30, pp. 57--77, 2015.

\bibitem{Xu2012b}
Q.~Xu, Q.~Huang, and Y.~Yao, ``Online crowdsourcing subjective image quality
  assessment,'' in \emph{ACM International Conference on Multimedia}, Nara,
  Japan, 2012, p. 359–368.

\bibitem{LIVE-IQA}
H.~Sheikh, M.~Sabir, and A.~Bovik, ``A statistical evaluation of recent full
  reference image quality assessment algorithms,'' \emph{IEEE Transactions on
  Image Processing}, vol.~15, no.~11, pp. 3440--3451, 2006.

\bibitem{Callet2005}
\BIBentryALTinterwordspacing
P.~Le~Callet and F.~Autrusseau, ``{Subjective quality assessment IRCCyN/IVC
  database}.'' [Online]. Available: \url{http://www.irccyn.ec-nantes.fr/ivcdb/}
\BIBentrySTDinterwordspacing

\bibitem{Qianqian2011}
Q.~Xu, T.~Jiang, Y.~Yao, Q.~Huang, B.~Yan, and W.~Lin, ``Random partial paired
  comparison for subjective video quality assessment via {HodgeRank},'' in
  \emph{ACM International Conference on Multimedia}, Scottsdale, Arizona, USA,
  2011, p. 393–402.

\bibitem{LIVE_VQA}
K.~Seshadrinathan, R.~Soundararajan, A.~C. Bovik, and L.~K. Cormack, ``Study of
  subjective and objective quality assessment of video,'' \emph{IEEE
  Transactions on Image Processing}, vol.~19, no.~6, pp. 1427--1441, 2010.

\end{thebibliography}


\end{document}